\documentclass[fleqn,twoside]{article}
\usepackage{espcrc2}
\usepackage{graphicx}

\usepackage{amssymb}

\newcommand{\ra}{\rightarrow}

\newcommand{\be}{\begin{equation}}
\newcommand{\ba}{\begin{eqnarray}}
\newcommand{\ea}{\end{eqnarray}}

\newcommand{\AmS}{{\protect\the\textfont2
  A\kern-.1667em\lower.5ex\hbox{M}\kern-.125emS}}

\title{
Photon and its hadronic interaction
}

\author{Maria Krawczyk\address[UW]{Institute of Theoretical Physics, Warsaw 
        University, \\
        ul. Ho\.za 69, 00-681 Warsaw, Poland}}

\begin{document}

\begin{abstract}
A short overview of  basics aspects of  hadronic interaction of the photon 
is presented.  
\end{abstract}
\maketitle

\section{Introduction}
Photon is one of the oldest elementary  particle with very well known (QED) 
properties, and as such it is considered to be a tool 
\underline{(``ideal probe'')} to test the structure of  more complicated 
objects like a proton.  
For example,  in the  DIS$_{ep}$, considered as the main source of information 
on the inner structure of the proton, the structure functions 
$F_{1,2,L}^{proton}$ for the proton can be measured. The probe - in a wide 
range of $Q^2$ - is provided by flux of virtual photons,  
emitted by  primary electrons (positrons). 
\begin{figure}[h]
\vskip -11.9cm\hskip -0.7cm
\includegraphics[scale=0.5]{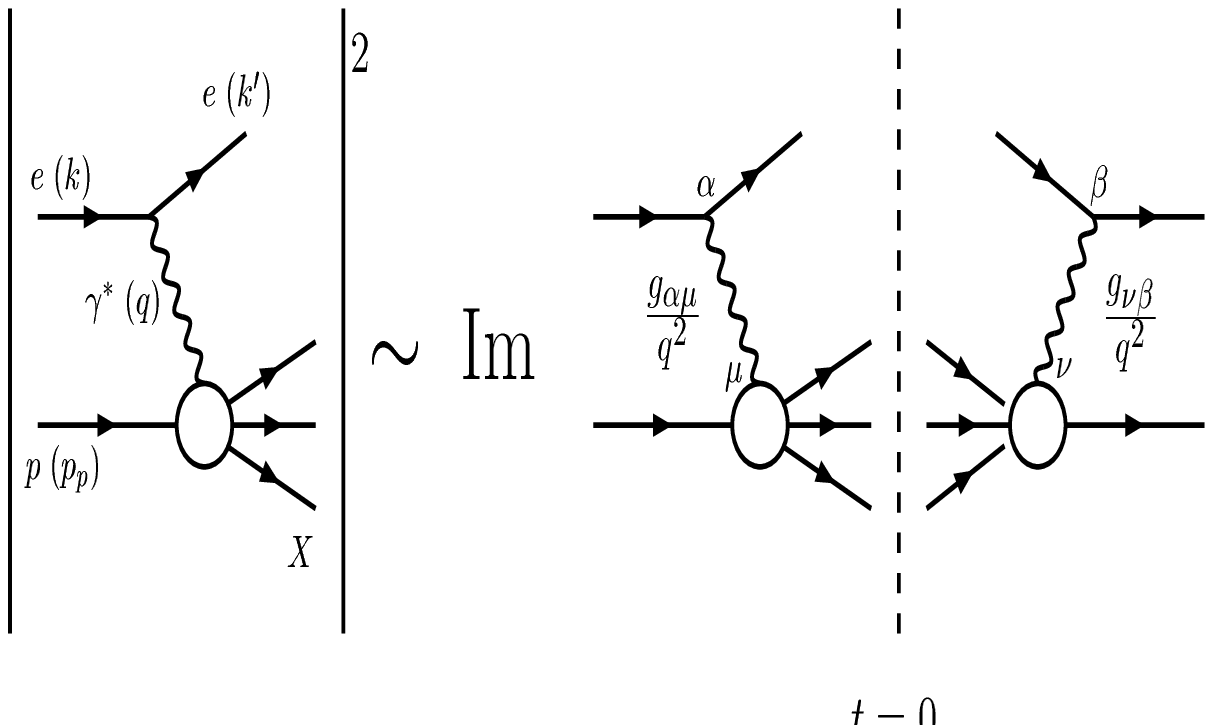}
\vskip -0.9cm
\end{figure}
So, besides proton also electron and  photon  are involved in 
the measurement of the structure functions and the parton densities of the 
proton. {A {\it factorization} allows to treat  independently
the electron  and proton part (figure above).  
The contributions of the longitudinal (L) and transverse (T) virtual  photons}
 can also be {\it separated}. Therefore, we can write
$$
d\sigma^{e p \rightarrow e X} \; = \;
\Gamma_T \; d\sigma^{\gamma^{\ast}_T p \rightarrow X} \; + \;
\Gamma_L \; d\sigma^{\gamma^{\ast}_L p \rightarrow X}. \; 
$$

Note, that the structure functions $F_{1,2,L}^{proton}$ are  proportional  to 
the combinations of the  cross sections for  processes 
$\gamma^*_{T,L} p \ra X$. One can ask if the photon is only a 
\underline{neutral probe} in such measurements and to what extend 
its apparent ``structure'' influences the corresponding structure functions
of the proton? 

Effects of the ``structure'' of the photon may influence our knowledge 
of the partonic content of the proton also in the large-$p_T$ photoproduction
at HERA.  However here both the initial (almost real) photon and 
the initial proton are on equal footing and it is natural nowdays
to take into account partonic content of both of them.

Similarly, the DIS events at HERA, if correspond to the large-$p_T$ single 
particle or jet production for $p_T^2\gg Q^2$, are sensitive to the 
``structure'' 
of the virtual photon (with virtuality $Q^2$). 

It is clear, that if we want to understand a proton, we should understand 
a photon itself. It is interesting to realized that approximately at the same 
time  (1922-3)  we have learned that the proton is not a fundamental particle,
while the photon is such a particle (Compton effect). 
Since early sixties XX we realized  that  the photon
in fact belongs to a ``hadron family'' with the closest relatives 
(according to the VMD idea)- the vector mesons:$\rho,  
\omega$ and  $\phi$. Next, the notion of a partonic content
of $\gamma$ have been introduced to describe in an effective way
its hadronic interaction.

\section{Photon}
At present colliders testing the  partonic content of the photon is a
difficult task, as at both LEP and HERA the photon is not  a primary particle.
The question if it is  simpler to deal with a partonic content 
of the electron was addressed during this workshop.

In the inclusive  process  $e^+e^-\ra e^+e^- hadrons$ one can probe the 
``structure'' of photon in various ways
depending on the  virtualities of emitted photons, 
{{$|q_1^2|=Q^2$ and  $|q_2^2|=P^2$}} (figure below): \\
 $\bullet$ The region {$Q^2\gg P^2 \gg \Lambda_{QCD}^2$ corresponds to 
{{DIS$_{e\gamma^*}$}}, where one can measure the 
structure functions of the  virtual photon.}\\
$\bullet$ For {$Q^2\gg \Lambda_{QCD}^2$, $P^2 \sim 0$ the corresponding 
process is
 {{DIS$_{e\gamma}$}},  where one can measure the 
structure functions of the real photon}.\\ 
$\bullet$ For {$Q^2\sim P^2 \sim 0$ the total cross section 
  {{$\sigma_{\gamma \gamma}^{tot}$}}} can be studied.
\begin{figure}[h]
\vskip -1.2cm,\hskip 1.2cm
\includegraphics[scale=0.35]{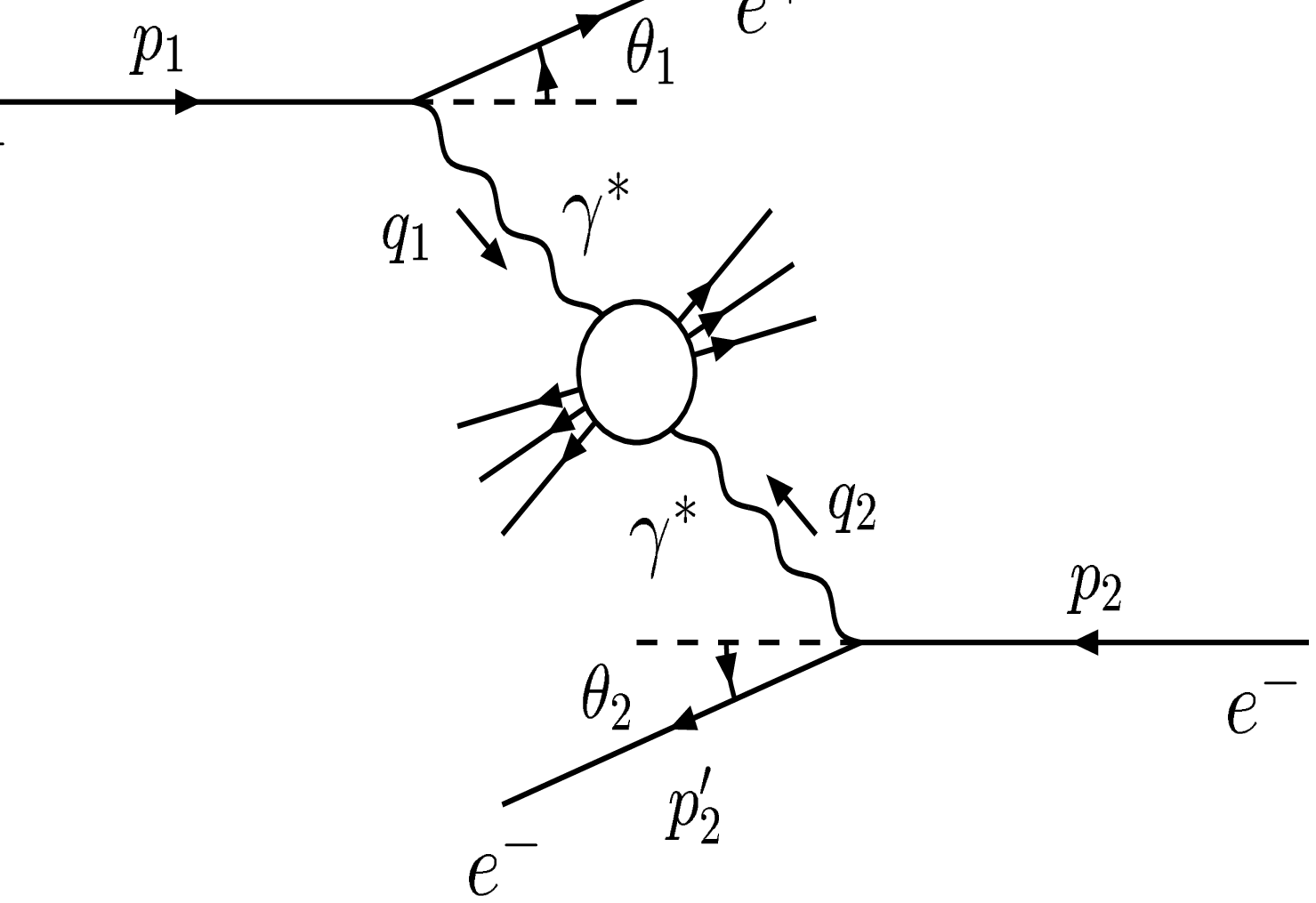}
\vskip -7.0cm
\end{figure}

In the DIS events $e \gamma \ra e X$   various structure functions for the 
real photons can be measured in principle,
however data are mainly for $F_2^{\gamma}$.

The $F_2^{\gamma}$  structure function  can be related to the  partonic 
densities of the 
real photon. In the Parton Model (PM) one obtains a  prediction from a pure 
QED that the quark densities  $q^{\gamma} \sim \ln Q^2$, and that $q^{\gamma}$ 
are large at the large value of $x$. In 
QCD there are leading $\ln Q^2$ corrections which can be sum up to all 
orders using an inhomogeneous DGLAP evolution 
equations. This form of equation allows to obtain an asymptotic solutions 
for quark densities in the photon without any   input. In practice,
one use an input usually based on  VMD (sometimes modelled by $\pi$ (sic!)).

In many analyses the  quark densities in $\gamma$ are treated as being
proportional to $\alpha/\alpha_s$, while in others to  $\alpha$, since
the main $\ln Q^2$-dependence mentioned above has nothing to do with QCD 
and $\alpha_s$. The way of counting has  practical consequences 
as it influences the choice  of  diagrams which define the NLO set.

The cross section for the  large-$p_T$ hadronic  processes involving  photons 
get contributions from  diagrams with the  direct  and the resolved photon
interaction with quarks. Note, that the resolved-photon  contribution are 
possible for real and virtual photons, in the latter case provided that
 $p_T^2$ is larger 
that $P^2$ - a ``virtuality'' of the $\gamma^*$.  Note that for the resolved
virtual photons processes one should take into account not only
cross section for a defined polarization of $\gamma^*$ (L or T) 
but also, since $\gamma$ is not a primary particle, the corresponding 
 interference terms: LT and T1 T2.

\section{New results for  proton and photon}
There many new data for the proton from HERA and for photon from the LEP 
(DIS and ''large-$p_T$'' events) and from HERA  (``large-$p_T$ events``), 
as presented at  this workshop.

There is  an impressive  theoretical progress  in the construction of the
parton parametrizations for the proton with a higher precision and using 
 various evolution equations, DGLAP, BFKL and  CCFM. Of a great importance are
special treatments of the heavy-quark thresholds and  error analyses of parton 
densities preformed recently. 
This  stimulates  the corresponding progress for a photon, 
reported during workshop.

\section{Old and new ideas}
\begin{itemize}
\item In 1934 L. de Broglie considered  ``The Neutrino Theory of Light'',
with the light quanta   composed of  $\nu \bar \nu$ pairs
(also Pauli, Heisenberg, Jordan, Kronig, Born). 
No  satisfactory derivation of Bose statistics of light quanta from 
Fermi statistics for $\nu$ was found and this idea was  abandoned (almost).
\vspace*{-0.3cm}
\item  The new idea is a Non-Commutative QED with a selfinteraction of photons.
\end{itemize}
\section{Instead of a summary}
{\it  The study of light has resulted in achievements of insight, 
imagination and ingenuity unsurpassed in any field of mental activity; 
it illustrates, too, better than any other branch of physics, 
the Vicissitudes of theories} ({Sir J.J. Thomson}, 1925).

For future study a Photon Collider with a real photon as 
a  primary particle will be very useful.
\vskip0.3cm
{\bf Acknowledgment:} I would like to thank the Local Organizing Committee
for this excellent workshop and all forms of support during my stay in 
Frascati. 
\end{document}